\documentclass[twocolumn,aps,prb,groupedaddress]{revtex4-1}
\usepackage{graphicx}
\usepackage{color}
\usepackage{amsmath}
\usepackage{enumitem}
\usepackage{amssymb}
\usepackage{hyperref}
\usepackage{svg}
\usepackage{ulem}

\newcommand{\be}{\begin{equation}}
\newcommand{\ee}{\end{equation}}

\newcommand{\bea}{\begin{eqnarray}}
\newcommand{\eea}{\end{eqnarray}}

\newcommand{\lp}{\left(}
\newcommand{\rp}{\right)}

\renewcommand{\Im}{{\rm \, Im\,}}

\renewcommand{\vec}[1]{{\boldsymbol #1}}

\newcommand{\addQ}[1]{\textcolor{magenta}{#1}}

\begin{document}
\title{
Activating superconductivity in a repulsive system by high-energy degrees of freedom
}


\author{Zhiyu Dong and Leonid Levitov} 
\affiliation{Massachusetts Institute of Technology, Cambridge, Massachusetts 02139, USA} 
\date{\today}

\begin{abstract}

We discuss superconductivity in a narrow conduction band sandwiched between unoccupied and occupied bands, an arrangement that enables an unconventional pairing mechanism governed by Coulomb repulsion. 
Pairing interaction originates from repulsion-assisted pair scattering between the conduction band and higher bands.
Optimizing the bandstructure design and 
carrier density in order to bring plasma frequency below the bandgap renders the interactions responsible for pairing unscreened, 
allowing superconductivity to fully benefit from the pristine Coulomb repulsion  strength. The repulsion-induced attraction 
is particularly strong in two dimensions and is assisted by low carriers density 
and plasma frequency values. We assess the possible connection of this mechanism to superconductivity in magic-angle twisted bilayer graphene where the bandstructure features wide dispersive upper and lower minibands. We use a simple model to illustrate the importance of the far-out pairs in these bands and predict testable signatures of this superconductivity mechanism. 
 
%
\end{abstract}

\maketitle


Achieving superconductivity using purely repulsive interactions has long been a topic of high interest.\cite{Kohn,Raghu2011,Chubukov_review,Lee2020} 
A well-established scenario for a repulsion-driven pairing is a two-stage framework in which Coulomb interactions create some lower-energy excitations, which, in turn, mediate carrier attraction near the Fermi level. This approach inspired the search for strong-coupling superconductivity in Hubbard models, in non-Fermi liquids and in the vicinity of quantum critical points.\cite{Abanov, Fradkin_review, Scalapino_review, Shibauchi_review}
It also helped to identify the unconventional pairing glue driving superconductivity in different systems, such as spin-density fluctuations in iron pnictides,\cite{Kamihara, Mazin1,Mazin2, Yao, Zhang2009a} plasmons\cite{Takada} in STO\cite{Ruhman1} and Bismuth,\cite{Ruhman2} and polaritons in quantum wells.\cite{Laussy} 

It is appealing, however, to consider an alternative route in which repulsive interactions generate attraction {\it directly,} rather than acting through low-energy intermediaries. 
A mechanism that achieves this goal would fully benefit from the strength of the pristine Coulomb interaction. 
One scenario of this type 
involves pairing between two or more Fermi surfaces 
\cite{Suhl, Geilikman, Aronov, Agterberg} 
or carrier pockets near 
van Hove (vH) points.\cite{Dzyaloshinskii, Schultz, Karyn, Lederer, Nandkishore, Fu_SC} However, in this mechanism 
the resulting pairing interaction is 
weakened by the Thomas-Fermi screening\cite{Geilikman, Aronov}, an effect that is particularly strong for narrow bands and near vH points.


\begin{figure}[t]
    \centering
    \includegraphics[width=0.45\textwidth]{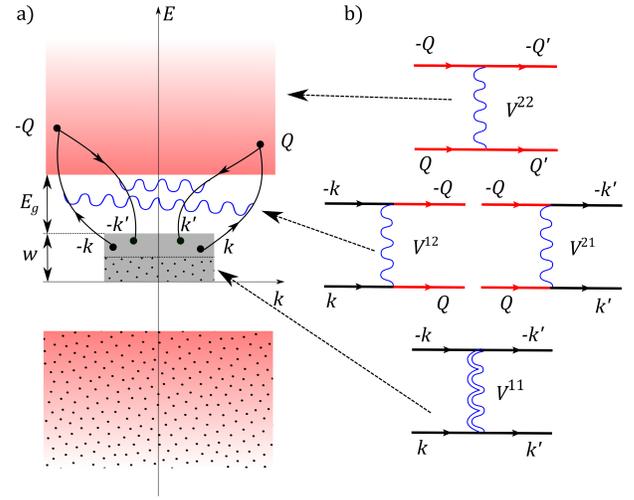}
    \caption{a) Schematic bandstructure and repulsion-assisted pair scattering that gives rise to an effective attraction. 
    The grey and red rectangles represent a narrow conduction band and a collection of filled and unfilled bands; 
dots indicate the occupied states. 
    The black arrows represent interband scattering:
    A pair state 
    ($k$,$-k$) near the Fermi level scatters to a large-momentum far-out pair ($Q$,$-Q$) in the upper band, and then back to the conduction band ($k'$,$-k'$). 
    b) Feynman diagrams for the four processes assisted by repulsion---the interband ($V^{12}$ and $V^{21}$) and intraband ($V^{11}$ and $V^{22}$) transitions. Repulsive interactions $V^{22}$ in the conduction band are subject to screening (indicated by a double wavy line) and are therefore small. Other interactions (single wavy lines) remain unscreened owing to a large frequency 
    transfer [see Eq.\eqref{eq:omega_p<E_g}]. 
    These processes dominate the effective pairing interaction.
    }
    \label{fig:two-band model simplified}
    \vspace{-5mm}
\end{figure}

In this article we describe a 
repulsion-based pairing mechanism which 
is not constrained by screening. 
Namely, we demonstrate an effective attraction in a purely repulsive system 
in a bandstructure with 
large-volume bands that are positioned just above and below the conduction band, as illustrated in Fig.\ref{fig:two-band model simplified}. 
The conduction band is narrow such that the plasma frequency is smaller than 
the band gap, 
\be\label{eq:omega_p<E_g}
 w_*< \omega_p < E_{\rm g}
,\quad w_*={\rm min}(E_F, w)
,
\ee
where $\omega_p$ is the plasmon mode frequency with momentum $k=k_F$ and $w_*$ is 
the pairing bandwidth in the conduction band in a one-band picture. 
Coulomb interaction creates interband pair scattering as illustrated in Fig.\ref{fig:two-band model simplified}. 
Under these conditions the unwanted effects of screening are limited to $\omega\lesssim\omega_p$.  This, by taking an advantage of an empty band position, 
generates a strong interband pair scattering. Unlike scenarios in which repulsion-assisted pairing originates from the frequency band $\omega<\omega_p$, in our mechanism the resulting pairing interaction near the Fermi level 
is not weakened by screening. 


Achieving strong pairing interaction 
mediated by the far-out pairs in the unoccupied bands
is facilitated by a small bandgap and high density of states in higher bands.
These effects can be optimized by tuning the bandstructure. 
Further, the direct Coulomb scattering in the conduction band is weakened by screening, an effect which is strong at $\omega<\omega_p$ but weak at higher $\omega$. As a result, the interaction at low energies is dominated by the interband processes shown in Fig.\ref{fig:two-band model simplified}, which give rise to an attraction. 
Therefore, screening helps our mechanism rather than hurts, presenting 
a considerable advantage over the two-Fermi-surface and vH pocket models.\cite{Suhl, Geilikman, Aronov, Agterberg, Nandkishore} 
The pairing interaction which is unaffected by screening 
fully benefits from the strength of the pristine Coulomb interaction.  

Further, 
the possibility of pairing in a narrow band coupled by pair scattering to the upper and lower bands, as shown in Fig.\ref{fig:two-band model simplified}, hints at interesting connections
to the 
superconductivity in magic-angle twisted bilayer graphene (MATBG)\cite{MacDonald,Cao1,Cao2}. From the start, the analogy with the phase diagram of high-$T_c$ superconductors\cite{Lee, Dagotto, Keimer} suggested 
that superconductivity in MATBG may originate from Coulomb repulsion, the strongest interaction in this system\cite{MacDonald,Cao1,Cao2,Fu_model, Fu_SC, Po_SC, Dmitri1, Dean}.
While we do not aim to 
settle this question here, 
the model of a conduction band with adjacent empty and filled bands is certainly MATBG-inspired. As such, it offers a useful perspective on achieving strong-coupling  
superconductivity due to purely repulsive interactions in tunable atomically thin nanosystems. Namely, it demonstrates that the high-energy states in an upper bands are capable of activating a strong-coupling superconductivity in a partially filled narrow band. 
This repulsion-driven pairing mechanism assisted by the far-out pairs has not been discussed in the field of MATBG superconductivity, where the theoretical and numerical study usually focuses on the flat bands from the start\cite{Girish, Cyprian_SC}.


This bandstructure offers useful knobs allowing to 
optimize the pairing interaction strength. 
Pair scattering to the upper and lower bands can be fully exploited by benefiting from a) low system dimension and narrow width $w$ of the conduction band, and b) large phase-space and slow band dispersion in the upper and lower bands.  
Here a) ensures that the Coulomb interaction is unscreened at large frequencies and is screened at $\omega\sim w<\omega_p$, whereas 
b) ensures that the contribution of the upper and lower bands to the pair susceptibility decays 
slowly with the increase of the momentum. 
This screens out the pair-breaking Coulomb interaction in the conduction band and enhances the contribution of the interband scattering to pairing interaction.  

\begin{figure}
    \centering
    \includegraphics[width=0.48\textwidth]{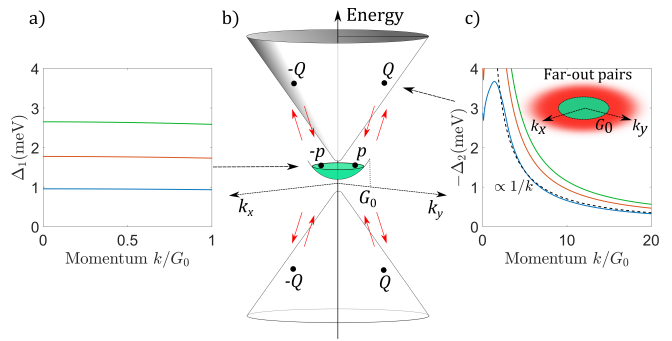}
    \caption{ 
    The two-band model and the predicted pairing amplitude in two bands. The middle panel 
    illustrates the bandstructure: 
  a narrow conduction band and higher bands. Shown are the pair states $\vec p$, $-\vec p$ near the Fermi level and $\vec Q$, $-\vec Q$ in the higher band and transitions between them (red arrows). Panels a) and c) show the pairing amplitude in each band, $\Delta_1$ and $\Delta_2$, obtained from Eq.\eqref{eq:BCS-self-consistency}. The curves correspond to different interband interaction values defined in Eq.\eqref{eq:g'}: $g'=400$ (blue), $g'=500$ (red) and $g'=600$ (green). The phase space for the pairs in the conduction band and the higher bands is illustrated in the inset of panel c). The phase volume for the far-out pairs in the higher bands is much larger than that for the pairs near the Fermi level. 
    }
    \label{fig:two-band model}
\end{figure}

Interestingly, these conditions are 
met by the MATBG bandstructure. 
The extremely high density of states in the MATBG's flat bands screens out the 
carrier repulsion in the conduction band \cite{Zachary}, whereas the interband pair scattering assisted by Coulomb interaction remains strong. 
The linear dispersion in the higher bands, together with the $1/q$ dispersion of Coulomb interaction in 2D, enhances the contribution of the far-out pairs. 
The relatively large phase volume of the higher bands ($\sim 10^4$ greater than the phase volume in the flat bands) helps to maximize the contribution of the far-out states to pairing. 
The sizeable pairing effect due to the far-out pairs is illustrated below in a simple model that mimics the MATBG bands [see Fig.\ref{fig:two-band model}]. 

Interestingly, recent measurements indicate that superconductivity in MATBG persists, and may even be slightly enhanced, in the presence of a close screening gate, whereas the correlated insulating order is suppressed.\cite{Stepanov,Liu} These findings are sometimes viewed as an evidence against a repulsion-based mechanism and in support of a phonon-mediated superconductivity. Yet, these findings can also be viewed as indicative of certain repulsion-assisted pairing mechanisms. Indeed, in 
the gated MATBG geometry\cite{Stepanov,Liu} only the 
harmonics of interaction with $kd\lesssim 1$ are screened out (here $d$ is the spacer width), whereas the harmonics $kd\gtrsim 1$ remain unscreened. 
As discussed below, the small-$k$ interactions give rise to pair breaking in the conduction band.  
To the contrary, the large-$k$ harmonics enable the far-out pairs that mediate pairing. 
Therefore, a mechanism relying on the far-out pairs predicts the dependence on gating in accord with the observations.\cite{Stepanov,Liu}  

To justify the scenario outlined above we 
consider a two-band model with the band 1 representing the narrow conduction band and the unoccupied band 2 being a proxy for the higher bands:
\begin{align}
    H= \sum _{i=1,2} \sum_{k} \varepsilon_{i\vec k} \psi^\dagger_{ik}\psi_{ik}
    .
    \label{eq:two-band-free-electron}
\end{align}
We take the states $\psi_{k}$ to be plane waves, the dispersion $\varepsilon_{i\vec k}$ is discussed below. 
%
%
To understand the 
repulsion-assisted pairing, we solve the self-consistency equation
\begin{equation}
    \Delta_i(\vec k) = -\int \frac{d\omega'}{2\pi}\sum_{i'=1,2} \frac{d^2\vec k'}{(2\pi)^2}  \frac{V^{ii'}_{\vec k \vec k'}\Delta_{i'}(\vec k')}{\omega'^2 +\varepsilon_{i'\vec k'}^2 +\Delta_{i'}^2(\vec k')}. \label{eq:BCS-self-consistency}
\end{equation}
The interaction $V^{ii'}_{\vec k \vec k'}$ describes scattering between pair states 
$\vec k,-\vec k$ and 
$\vec k',-\vec k'$ in bands $i$ and $i'$; the electron energies are measured from the Fermi level. 
We take the interaction as dynamically screened Coulomb repulsion: 
\be\label{eq:V^ij_toy_model}
    V^{ii'}_{\vec k \vec k'} = \sum_q 
    V(q,\delta\omega^{ii'}) |A^{ii'}_{\vec k \vec k',\vec q}|^2
    \approx \frac{V_0(|\vec k-\vec k'|)}{\epsilon(\vec k-\vec k',\delta\omega^{ii'})}. 
\ee
Here $\sum_q$ denotes $\int \frac{d^2q}{(2\pi)^2}$, $V_0(k)=\frac{2\pi e^2}{k}$ is the bare interaction, 
$\delta\omega^{ii'}$ is the frequency change in the scattering. 
For simplicity, the overlaps between different Bloch functions 
$A^{ii'}_{\vec k \vec k',\vec q} \equiv \langle u^i_{\vec k} |e^{i(\vec k-\vec k'+\vec q) \vec r} | u^{i'}_{\vec k'} \rangle$ are taken to be unity. 

Crucially, the frequency dependence originating from dynamical screening can result in strong suppression of the pair-breaking intraband processes $V^{11}$ compared to the interband processes $V^{12}$ and $V^{21}$ that facilitate pairing. This regime, favorable for pairing, is achieved when the plasma frequency $\omega_p$ is greater than the width of the conduction band but smaller than the bandgap $E_{\rm g}$. 
While the intraband processes $V^{11}$ are strongly screened, the interband processes $V^{12}$ and $V^{21}$ are screened only by the interband polarization, yielding values 
$V^{12}$ and $V^{21}$ at least an order of magnitude greater that $V^{11}$ [see Supplement \ref{sec:supplement_interaction}].
This allows the pairing interaction due to interband processes to dominate over the pair-breaking effect due to the intraband processes. 


The two-band pairing problem, 
Eq.\eqref{eq:BCS-self-consistency}, has a straightforward solution which is described below.
However, in order to illustrate the essential physics, it is instructive to first inspect a simple toy model. 
In that, we ignore the momentum and frequency dependence of the 
repulsive interactions $V^{ii'}_{\vec k \vec k'}$, retaining only the dependence on the band index $i$, $i'$, with all values being positive: 
    \begin{align}
        V^{ii'}_{\vec k \vec k'} = V^{ii'}>0 
        .
    \end{align}
In this case $\Delta_i(\vec k)$ has no $\vec k$ dependence and
the gap equation 
assumes the form of two coupled 
algebraic equations:
\begin{align}
    &  
    \begin{bmatrix}
    \Delta_1\\
    \Delta_2
    \end{bmatrix} = 
     \begin{bmatrix}
    -V^{11} Z_1 & -V^{21} Z_2\\
    -V^{12} Z_1 & -V^{22} Z_2
    \end{bmatrix}
    \begin{bmatrix}
    \Delta_1\\
    \Delta_2
    \end{bmatrix}. \label{eq:two-sector-toy-model-self-consistency}
\end{align}
The quantities $Z_1$ and $Z_2$ 
denote the Cooper pair susceptibilities in each band, 
defined as
\begin{align}
    Z_1 &= \int \frac{d\omega'}{(2\pi)^3} \int_{0}^{G_0} \frac{d^2\vec k'}{\omega'^2 +\varepsilon_{1\vec k'}^2 +\Delta_{1}^2} = N(0) \ln \frac{w}{\Delta_1}, \label{eq:Z_1} 
    \\
    Z_2 &= \int \frac{d\omega'}{(2\pi)^3}\int_{0}^{G_0} \frac{d^2\vec k'}{\omega'^2 +\varepsilon_{2k'}^2 +\Delta_2^2}\sim \frac{G_0^2}{ E_{\rm g}}. \label{eq:Z_2}
\end{align}
Here $G_0=2\pi/a_0$ is the Brillouin zone size, $a_0$ is the superlattice period, $w$ is the conduction band width. 
We model the density of states in the conduction band as a constant, of value $N(0)$
per one spin projection. For illustration, we take the higher band to be dispersionless, positioned at the energy $E_g$ away from the Fermi level.  
%
%



The two-band problem in Eq.\eqref{eq:two-sector-toy-model-self-consistency} yields a gap equation
\begin{equation}
    1 = \tilde{g}N(0)\ln\frac{w}{\Delta_1},
\end{equation}
with a net effective interaction value 
\begin{equation}
    \tilde{g} = -V^{11}+ 
    \frac{V^{12}V^{21} Z_2}{1+V^{22}Z_2}. \label{eq:two-band-result}
\end{equation}
Since the values $V^{ij}$ are positive for a repulsive interaction, a positive value $\tilde{g}$ means a net effective attraction that facilitates pairing.


\addQ{}
We consider the case where $Z_2$ is large (which can be realized either by making the $E_{\rm g}$ small or by introducing more upper-band states) whereas the several couplings associated with the upper-band states are strong and of comparable strength, i.e.
\begin{equation}
    V^{12}Z_2 \gg 1, \quad V^{12}\sim V^{21} \sim V^{22}.
\end{equation}
In this limit the effective pairing interaction becomes 
\begin{equation}
    \tilde{g} \approx  -V^{11} + \frac{V^{12}V^{21}}{V^{22}}. \label{eq:effective_interaction}
\end{equation}
This result confirms the picture discussed above. The first term, which for a repulsive interaction describes pair breaking, corresponds to the intraband processes with a small frequency transfer, whereas the three couplings in the second term are associated with the states in the upper band corresponding to the processes with a large frequency change. The dynamical screening will suppress the first term but will have little effect on the second term. Therefore, the main effect of the dynamical screening is to suppress pair breaking and enhance the effective pairing interaction.

Next, we upgrade the toy model to a more realistic model, accounting for the dispersion of higher bands and for the interaction $\vec k$ dependence. The higher minibands in the MATBG bandstructure, if viewed in extended zone scheme, can be approximately represented as two copies of graphene monolayer Dirac bands (one per each layer):
\begin{align}
    \varepsilon_2(\vec k) = \pm \sqrt{v^2|\vec k|^2+ E_{\rm g}^2}, \quad |\vec k|<G 
    ,
    \label{eq:E_2(k)_MATBG_}
\end{align}
with momentum taking values 
in the monolayer graphene Brillouin zone 
(see Fig.\ref{fig:two-band model}). The exact value of the bandgap 
$E_{\rm g}$ (roughly $20$ meV in MATBG) 
is of marginal importance for our analysis, since it only affects the pair susceptibility in a region near the bottom of the upper band, small compared to the large phase volume of the far-out pairs.
Here we use $G/G_0 \sim 10^{2}$, a value that mimics MATBG superlattice parameters.



Next, we consider the interactions. 
As argued above, strong screening in the conduction band is crucial since it suppresses the pair breaking interaction, leaving 
the (useful) interband processes unaffected. 
To confirm that this is indeed the case, we estimate the Thomas-Fermi screening in the conduction band [modelled as above, see Supplement], arriving at a relatively small value
\begin{align}
    V^{11}_{\vec k \vec k'} \sim V(q<G_0) \sim 100 \text{ meV nm}^2. \label{eq:V^11_}
\end{align}
In comparison, pair scattering processes into and within higher bands remain
unscreened because the frequency transfer in such processes is larger than plasma frequency. 
These interactions approximately follows the form of the 
bare interaction $\sim 1/k$.
After accounting for interband polarization and umklapp scattering (see Supplement \ref{sec:supplement_interaction}), the interactions $V^{12}$,  $V^{21}$ and  $V^{22}$ 
can be modeled as a $1/k$ potential with a small-$k$ cutoff at 
$k<k_*$:
\be
    V^{12}_{\vec k \vec k'} =V^{21}_{\vec k \vec k'}=  V^{22}_{\vec k \vec k'} =  \frac{g'G_0}{k_*+|\vec k-\vec k'|},\quad g' = \frac{2\pi e^2}{\epsilon G_0}
    . \label{eq:V^12V^21V^22_new_}
\ee
The cutoff value $k_*$, 
originating from umklapp scattering, is set by the inverse of the Wannier orbital radius. We use $k_* = 2G_0$ in our numerical calculation for simplicity. Estimating 
$g’$ we find a value 
$\sim 10$ times larger than the value of $g$ in Eq.\eqref{eq:V^11_} [see Supplement \ref{sec:supplement_interaction}]. 
Using the quantities given above, 
we solve the gap equation, Eq.\eqref{eq:BCS-self-consistency}, to obtain 
pairing amplitudes in the two bands 
[see Fig.\ref{fig:two-band model}]. 

The amplitude $\Delta_1$ is nearly $k$-independent, a result consistent with the picture  
of far-out pairs generating a constant effective attraction for the states 
in the conduction band. Notably, the size of $\Delta_1$ obtained in this way is comparable to  $T_c$ measured in MATBG. In contrast, the pairing amplitude in the higher bands 
strongly depends on momentum, behaving as
$\Delta_2\sim 1/k$ over a wide range of $k$
(see Fig.\ref{fig:two-band model}). This behavior is due 
to the $1/k$ dependence of 
interaction at large $k$. 
A direct outcome of repulsive interactions is that $\Delta_1(k)$ and $\Delta_2(k)$ have opposite signs, 
similar to the case of $s_{\pm}$ 
order in iron pnictides\cite{Mazin1,Mazin2} where the pairing amplitudes have opposite signs in different Fermi pockets. 



\begin{figure}
    \centering
    \includegraphics[width=0.5\textwidth]{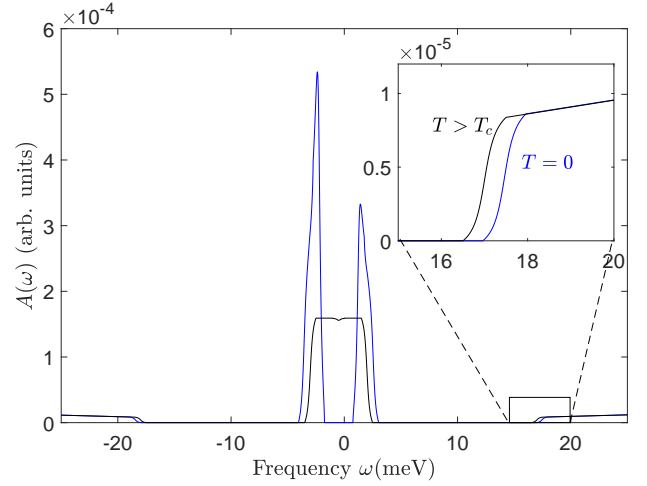}
    \caption{Schematic of the local spectral function in the normal state (black) and the superconducting state (blue). Shown is $A(\omega)= \sum _{\vec k} A(\vec k,\omega)$ with $A(\vec k, \omega)$ given in Eq.\eqref{eq:spectral_function}, calculated for $g'=500$ \rm{meV nm$^2$}. 
    A signature of the far-out pairs and finite pairing amplitude $\Delta_2$ in the higher bands is a shift of the minimal frequency of the higher-band contribution appearing simultaneously with gap opening at the Fermi level in the conduction band. 
    }
    \label{fig:spectral_function}
\end{figure}

Interestingly, 
the pairing amplitude value in the higher band is comparable or even larger than the one in the conduction band. 
The large $\Delta_2$ manifests itself in observables that are directly sensitive to the pair amplitude in higher bands. 
One such observable is 
the spectral function $A(\vec k,\omega) = \frac{1}{\pi}\Im \sum_{ \alpha} G_\alpha(\vec k,\omega^-)$. 
In the superconducting phase this quantity is expressed through the quasiparticle spectrum and coherence factors as
\be\label{eq:spectral_function}
A(\vec k,\omega) = \sum_{\alpha} u_{\vec k\alpha}^2\delta(\omega-E_{\vec k \alpha}) + v_{\vec k\alpha}^2\delta(\omega+E_{\vec k\alpha})
\ee
with $\alpha=1,2$ the band index, $\omega^-= \omega-i0^+$, and
\[
u_{\vec k\alpha}^2, v_{\vec k\alpha}^2 = \frac1{2}\pm \frac{\varepsilon_{\vec k\alpha}}{2E_{\vec k\alpha}}
,\quad E_{\vec k\alpha} = \sqrt{\varepsilon_{\vec k\alpha}^2+\Delta_{\alpha}(\vec k,\omega)^2}
.
\]
The superconducting order changes the quasiparticle spectrum 
and reshapes the spectral function as illustrated in Fig.\ref{fig:spectral_function}. One signature of pairing is the $2\Delta_1$ gap opening at the Fermi level, 
another is a shift of band edges by an amount that depends on the $\Delta_i$ value. 
The shift is $\sqrt{\epsilon_{\rm 1,max}^2+\Delta_1^2}-\epsilon_{\rm 1,max}$ for the conduction band and $\sqrt{\epsilon_{\rm 2,min}^2+\Delta_2^2}
-\epsilon_{\rm 2,min}$ for the higher band. 
For $g'$ values used in Fig.\ref{fig:two-band model}, $\Delta_2$ can be as large as few meV near the bottom of the higher band. As illustrated in Fig.\ref{fig:spectral_function}, this can produce a measurable energy shift. 

Spectral functions can be probed using the direct-space scanning tunneling spectroscopy\cite{Binnig, Chen_book, Stroscio_book, Wiesendanger_book, Hoffman_review}, the Fourier-transformed scanning tunneling spectroscopy\cite{Crommie,Hasegawa,Wittneven,Burgi,Fujita,Kanisawa,Hoffman,Zhang2009b,Roushan} and other state-of-the-art techniques. 
The differences in the spectral function between the superconducting and normal states, 
in particular the features associated with the higher bands, can provide direct evidence for the far-out pairs 
and the repulsion-assisted 
superconductivity. 



As discussed above, recent observations of superconductivity becoming stronger under close gating\cite{Stepanov,Liu} lend support to the scenario outlined above. 
A proximal gate which is not too close to MATBG primarily screens out the $V^{11}$ interactions but has little effect on the $V^{12}$, $V^{21}$ and $V^{22}$ interactions that enable the far-out pairing mechanism. Therefore, since $V^{11}$ is the pair-breaking interaction, superconductivity is strengthened by close gating. To the contrary, a phonon-mediated superconductivity is expected to remain unaffected by gating or to become weaker due to screening of the deformation potential, similar to the observed extinction of the correlated insulating states\cite{Stepanov,Liu}. It is of course possible that phonons and a repulsion-assisted mechanism act simultaneously,  
in this case a more comprehensive study 
will be required to determine the dominant pairing mechanism.

Very recently superconductivity was discovered in twisted trilayer graphene\cite{Park-Cao, Hao} where the band structure is similar to MATBG. Further investigation\cite{Cao-Park} indicates the spin-triplet pairing order. Since achieving spin-triplet order through a phonon mechanism is unlikely, these findings add evidence for a Coulomb repulsion-based pairing mechanism. 

As a cautionary remark, despite the prediction of a strong pairing effect being of obvious interest, pinning the explanation of SC in MATBG on this mechanism would be premature. 
More work will be needed to compare this mechanism 
with other possible pairing glues. 
However, regardless of whether this mechanism 
explains the superconductivity in MATBG, it certainly plays a non-negligible role in this system. Further, it give rises to a unique effect --- pairing in the upper bands, which in itself represents an interesting direction for future work. 

This work benefited from useful discussions with Dmitri Efetov, Fransisco Guinea, Pablo Jarillo-Herrero and Patrick Lee.

\newpage

\widetext
\begin{center}
\textbf{\large Supplementary Information}
\end{center}
\section{The interaction parameters in two-band model}\label{sec:supplement_interaction}
In this section, we estimate the interaction $g$ and $g'$ in the two-band model used in main text. 
As a reminder, the scattering amplitude takes the following form:
\begin{equation}
    V^{ii'}_{\vec k \vec k'} = \frac{V_0(|\vec k-\vec k'|)}{\epsilon(\vec k-\vec k',\delta\omega^{ii'})}, \quad V_0(q)=\frac{2\pi e^2}{q}.\label{eq:V^ij}
\end{equation}
where $\epsilon$ is the dielectric constant, the $\delta\omega^{ii'}$ is the change of frequency in the scattering process. Here, we have taken the Bloch function as plane waves for simplicity.  
Below, we discuss the value of $V^{ii'}_{\vec k\vec k'}$ for different $i$ and $i'$.  

For scattering inside the flat band, i.e. $V^{11}_{\vec k \vec k'}$, the pairs involved have a characteristic frequency comparable to the width of the flat band $w$. As discussed in the main text, the frequency transfer in the intraband scattering is lower than the plasma frequency.
The corresponding value $V^{11}$ can be estimated by replacing the dynamical one in Eq.\eqref{eq:V^ij} with the static dielectric function $\epsilon_0(q) = 1+V_0(q)\Pi_0(q)$. 
We estimate polarization function $\Pi_0(q)$ using the density of states in MATBG. Given the bandwidth $\sim 5\rm{meV}$ and the size of superlattice unit cells $\sim 10^2\rm{nm}^2$, the density of state in each spin-valley is approximately $N(0)\sim 2\times 10^{-3} \rm{meV}^{-1} \rm{nm}^{-2}$. Accounting for the four degenerate spin-valley species which all contribute to the screening, we estimate the polarization as 
\be
\Pi_0(q<G_0)\sim 4N(0) \sim 1\times 10^{-2} \rm{meV}^{-1}\rm{nm}^{-2}.
\ee
As a result, the strength of the intraband repulsion is:
\begin{align}
    V^{11}_{\vec k \vec k'} \sim V(q<G_0) \sim \frac{1}{\Pi_0(q)} \sim 100 \text{ meV nm}^2. \label{eq:V^11}
\end{align}
where in the second step we have used $V_0(q)\Pi_0(q)\gg 1$.

Next, we consider the amplitudes of the processes that contain states in higher bands. In these processes, the momentum transferred in the scattering is usually large ($q>G_0$). At such large momentum, the screening effect is dominated by the interband polarization which transfers a frequency larger than plasma frequency. 
Due to the close relation between dispersive band and the Dirac band in monolayer graphene, the interband polarization can be estimated as the polarization in two sheets of pristine graphene. 
Therefore, we use the value of two times the dielectric constant in monolayer graphene (MLG) as the dielectric constant of the large-momentum scattering processes in MATBG: $\epsilon(q>G_0) \sim 2\epsilon_{MLG} \sim 10$. As a result, the interaction at large momentum is given by the following form:
\begin{align}
    & V_>(q) \equiv V(q>G_0) = g' \frac{G_0}{q}, \label{eq:V(q>G_0)}\\
    & g' = \frac{2\pi e^2}{\epsilon G_0} \sim 1\times 10^3 \text{meV nm}^2. \label{eq:g'}
\end{align}
If we further account for the dielectric constant of the substrate, the dielectric constant will be sightly larger. Accordingly, in numerical calculation presented in Fig.\ref{fig:two-band model}, we use slightly smaller $g'$: $g'\sim 500$meV nm$^2$. The three scattering amplitudes in Eq.\eqref{eq:V^ij} that contain upper-band states can be written as
\begin{align}
    V^{12}_{\vec k \vec k'} =V^{21}_{\vec k \vec k'}=  V^{22}_{\vec k \vec k'} = g' \frac{G_0}{|\vec k- \vec k'|}.
    \label{eq:V^12V^21V^22}
\end{align}
These amplitudes are stronger than the intraband interaction $V^{11}$.

We emphasize that the $1/|k-k'|$ dependence in interband scattering should be cut off below some finite momentum $|k-k'|\sim k_*$ due to the umklapp scattering. The reason is that the flatband wavefunction $\psi_k$ contains multiple plane wave components: $\psi_{1\vec k}(\vec r) =  \sum_{\vec n }c_{1,\vec k,\vec n}e^{i(\vec k+\vec G)\vec r}$, where $\vec n = (n_x,n_y)$, $n_x, n_y \in \mathbb{Z}$. 
Such cut off is imposed through the following form:
\be\label{eq:V12_V21_sup}
V^{12}_{\vec k \vec k'} =V^{21}_{\vec k \vec k'} = g' \frac{G_0}{k_*+|\vec k- \vec k'|}.
\ee
In numerical analysis discussed in main text, we have set $k_* = 2G_0$. 

In contrast, $V^{22}$ is free from the cutoff since the states in the dispersive band resemble the Bloch wavefunction in monolayer graphene and thus only contain a single-plane wave component. So $V^{22}$ strictly follows the $1/k$ dependence:
\be\label{eq:V22_sup}
V^{22}_{\vec k \vec k'} = g' \frac{G_0}{|\vec k- \vec k'|}.
\ee
The scattering amplitudes introduced in Eq.\eqref{eq:V12_V21_sup} and Eq\eqref{eq:V22_sup} is a heuristic way to mimic the realistic MATBG system. Using this model, we avoid the complication of real Bloch wavefunction in MATBG, while qualitative capturing the momentum dependence in the scattering matrix. 

\section{The two-band gap equation with a separable pairing interaction}\label{sec:supplement:separable_interaction}
In this section, we analyze the gap equation using a soluble model.
In order to solve the gap equation analytically while maximally capturing the momentum dependence described in Eq.\eqref{eq:V^11}, Eq.\eqref{eq:V12_V21_sup} and Eq.\eqref{eq:V22_sup}, we rewrite the $V^{ii'}_{kk'}$ into the following separable form:
\begin{align}
    V^{ii'}_{\vec k \vec k'} = 
    g \delta_{i1}\delta_{i'1} +
    \frac{g' G_0}{ \sqrt{kk'} } \lp 1-\delta_{i1}\delta_{i'1} \rp.
\end{align}
The pairing  
in this two-band system $\Delta_i(k)\sim\langle \psi_{ik}\psi_{ik} \rangle$ inherits the structure of the scattering amplitudes:
\begin{align}
    \Delta_i(k) = \Delta_{1} \delta_{i1} + \Delta_{2}\sqrt{\frac{G_0}{k}} \delta_{i2}.
\end{align}
This yields the following self-consistency equations:
\begin{align}
    \Delta_{1} &= -\frac{1}{2}g \int_0^{G_0}\frac{d^2\vec k'}{(2\pi)^2} \frac{\Delta_1}{\sqrt{\varepsilon_1^2(k') +\Delta_1^2}} -\frac{1}{2} Ng' \int_{G_0}^{G}\frac{d^2\vec k'}{(2\pi)^2} \frac{G_0}{vk'^{2}} \Delta_2, \label{eq:two-sector-Delta<}\\
    \Delta_{2} &= -\frac{1}{2}g' \int_0^{G_0}\frac{d^2\vec k'}{(2\pi)^2}\sqrt{\frac{G_0}{k'}} \frac{\Delta_1}{\sqrt{\varepsilon_1^2(k') +\Delta_1^2}}  - \frac{1}{2} Ng' \int_{G_0}^{G}\frac{d^2\vec k'}{(2\pi)^2} \frac{G_0}{vk'^2}\Delta_2. \label{eq:two-sector-Delta>}
\end{align}
where $N=4$ is the degeneracy of Dirac band in MATBG (two layers times particle/hole Dirac bands). In the last term in the first line, we have taken $k\sim G_0$, i.e. the ``small momentum" of interest is at the order of $G_0$. 
Working out the integrals we get:

\bea\label{eq:self-consistency-final}
    \Delta_{1} &= - g N(0) \Delta_1 \ln \frac{2w}{\Delta_1}  - \frac{1}{2} g'  \frac{NG_0}{2\pi v} \ln{\frac{G}{G_0}} \Delta_{2},  \nonumber\\
    \Delta_{2} &= - g' N(0) \Delta_1 \ln \frac{2w}{\Delta_1} - \frac{1}{2} g' \frac{NG_0}{2\pi v} \ln{\frac{G}{G_0}} \Delta_2.
\eea
where $N(0)$ is the density of states for one of the four flavors in MATBG (two spins times two valleys), $v\sim 10^6\mathrm{m/s}$ is the velocity of the dispersive band, i.e. the velocity of the Dirac band in the monolayer graphene.
The solution is given by:
\begin{align}
    1 &= \tilde{g} N(0) \ln \frac{2w}{\Delta_1},
\end{align}
where $\tilde{g}$ is interpreted as the effective pairing potential, which is given by
\begin{equation}
    \tilde{g} = -g + \Delta g, \quad \Delta g = \frac{ g'^2 Z_2}{1+ g' Z_2},\quad Z_2 = \frac{NG_0}{4\pi v} \ln\frac{G}{G_0}. \label{eq:effective_interaction_MATBG}
\end{equation}
Finally, we plug in the numerical values 
and found that the attraction due to higher bands is:
\begin{equation}
    \Delta g = 590 \text{ meV nm}^2,
\end{equation}
which is larger than the intra-band repulsion $g$. 
This result confirms that a net attraction can be generated in MATBG from the far-out pair mechanism. 

\end{document}